\documentclass[preprint,aps]{revtex4}
\usepackage{graphicx}
\usepackage{dcolumn}
\usepackage{bm}
\usepackage{multirow}
\usepackage{color}
\usepackage[colorlinks=true]{hyperref} 
\usepackage{amsmath}
\usepackage{amssymb}
\usepackage{epsfig}
\usepackage[T2A]{fontenc}
\usepackage[cp1251]{inputenc}
\usepackage[russian,english]{babel}
\usepackage{array}
\usepackage{gensymb}

\begin{document}

\title{Electrical initialization of electron and nuclear spins in a single quantum dot at zero magnetic field}

\author{F. Cadiz$^1$}
\author{A. Djeffal$^2$}
\author{D. Lagarde$^1$}
\author{A. Balocchi$^1$}
\author{B. S. Tao$^{2,5}$}
\author{B. Xu$^4$}
\author{S.H. Liang$^2$}
\author{M. Stoffel$^2$}
\author{X. Devaux$^2$}
\author{H. Jaffres$^3$}
\author{J.M. George$^3$}
\author{M. Hehn$^2$}
\author{S. Mangin$^2$}
\author{H. Carrere$^1$}
\author{X. Marie$^1$}
\author{T. Amand$^1$}
\author{X. F. Han$^5$}
\author{Z. G. Wang$^4$}
\author{B. Urbaszek$^1$}
\email{urbaszek@insa-toulouse.fr;yuan.lu@univ-lorraine.fr;renucci@insa-toulouse.fr}
\author{Y. Lu$^2$}
\author{P. Renucci$^1$}

\affiliation{$^1$Universit\'e de Toulouse, INSA-CNRS-UPS, LPCNO, 135 Avenue Rangueil, 31077 Toulouse, France}
\affiliation{$^2$Institut Jean Lamour, UMR 7198, CNRS-Universit\'e  de Lorraine, Campus ARTEM, 2 Allee Andre Guinier, BP 50840, 54011 Nancy, France}
\affiliation{$^3$Unit\'e Mixte de Physique CNRS/Thales and Universit\'e Paris-Sud, 1 Avenue Augustin Fresnel, 91767 Palaiseau, France}
\affiliation{$^4$Key Laboratory of Semiconductor Materials Science, Institute of Semiconductors,
Chinese Academy of Sciences, P. O. Box 912, Beijing 100083, China}
\affiliation{$^5$Beijing National Laboratory of Condensed Matter Physics, Institute of Physics, Chinese Academy of Sciences, Beijing 100190, China}

\begin{abstract}
The emission of circularly polarized light from a single quantum dot relies on the injection of carriers with well-defined spin polarization. Here we demonstrate single dot electroluminescence (EL) with a circular polarization degree up to 35\% at zero applied magnetic field. The injection of spin polarized electrons is achieved by combining ultrathin CoFeB electrodes on top of a spin-LED device with p-type InGaAs quantum dots in the active region. We measure an Overhauser shift of several $\mu$eV at zero magnetic field for the positively charged exciton (trion X$^+$) EL emission, which changes sign as we reverse the injected electron spin orientation. This is a signature of dynamic polarization of the nuclear spins in the quantum dot induced by the hyperfine interaction with the electrically injected electron spin. This study paves the way for electrical control of nuclear spin polarization in a single quantum dot without any external magnetic field. 
\end{abstract}


\maketitle

\textbf{Introduction.---} 
Efficient electrical spin injection into semiconductors is the prerequisite to operating any spintronic or spin-based quantum-computing scheme using semiconductors. Spin Light Emitting Diodes (SpinLEDs) \cite{Fiederling:1999a,Ohno:1999a,Nishizawa:2017a,Hanbicki:2003a} allow efficiently generating and detecting spin polarized currents up to room temperature \cite{Jiang:2005a} in a  semiconducting active region. Here a key issue is the injection in the so-called tunnel regime. This allows circumventing the conductivity mismatch problem between the ferromagnetic (FM) electrode and the semiconductor \cite{Fert:2001a} by introducing MgO tunnel barriers \cite{Jiang:2005a,Lu:2008a,Tao:2016a}. Optically active semiconductor nanostructures such as quantum dots are excellent model systems for various applications \cite{Salter:2010a,Santis:2017a}: compact sources (Spin LEDs\cite{Li:2005a,Itskos:2006a,Kioseoglou:2008a,Merz:2014a}, Spin Lasers\cite{Holub:2007a}) of polarized light (for information science\cite{Farshchi:2011a}, detection of chirality in life science\cite{Xu:2006a}, 3D screens\cite{Kim:2006a}) based on p-i-n junctions, as well as single quantum bits for quantum computation \cite{Henneberger:2016a}.\\
\begin{figure*}
\includegraphics[width=0.95\linewidth]{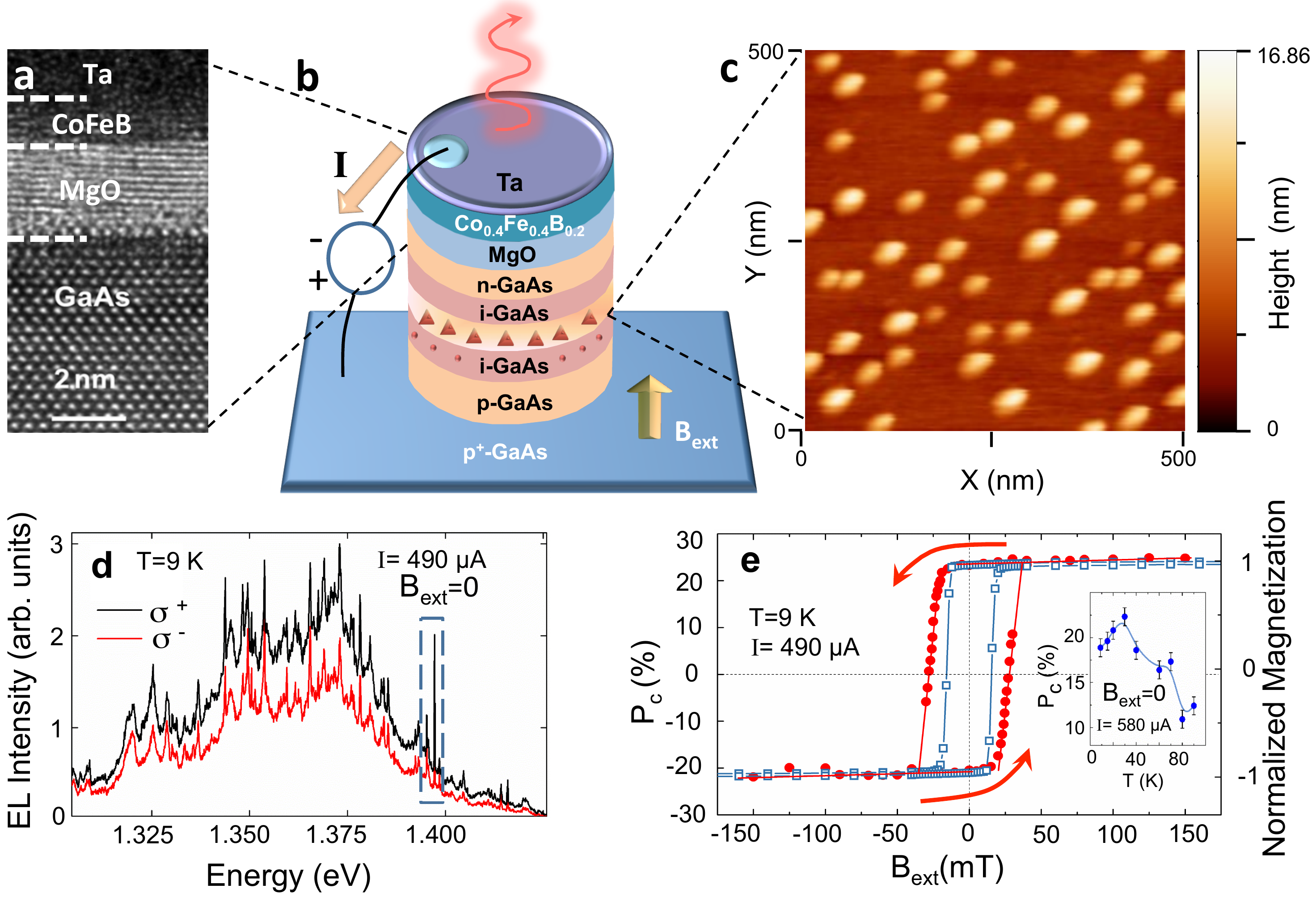}
\caption{\label{fig:fig1} \textbf{Spin LED device with p-doped InAs/GaAs quantum dots and polarization resolved electroluminescence of an ensemble of quantum dots.}    \textbf{a.} High Resolution-Transmission Electron Microscope image of the injector Ta/CoFeB/MgO/GaAs. \textbf{b.} Schematic structure of the spin-LED device. A single layer of InAs QDs is embedded in the intrinsic region of the p-i-n junction of the LED.  \textbf{c. }AFM image of InAs QDs with a density of $1.6\times 10^{14}~m^{-2}$. The average lateral dot diameter is about 30~nm and the height 9~nm. \textbf{d.} Electroluminescence from the device shows spectrally narrow emission lines stemming from an ensemble of semiconductor quantum dots. The applied magnetic field for the measurement is zero, and the magnetization of the CoFeB layer has been saturated before the measurements. $\sigma^+ (\sigma^-)$ polarized EL signal is plotted in black (red). \textbf{e.} The circular polarization degree of the EL (red circles) is plotted as a function of applied magnetic field $B_{ext}$ for the ensemble emission of panel d. Hysteresis loop of the normalized magnetization of CoFeB electrode measured by SQUID at T=30~K (blue squares). The inset shows the evolution of $P_c$ with temperature at $B_{ext} = 0$.}
\end{figure*}
\indent 
So far, the generation of spin polarized carriers in a single quantum dot required either optical pumping with circularly polarized lasers or the application of an external magnetic field (of several Tesla) for devices based on electrical spin injection from a magnetic electrode \cite{Loffler:2007a,Ghali:2008a,Kioseoglou:2008a,Asshoff:2011a,Asshoff:2011b}, which is not convenient for practical applications. For circularly polarized electroluminescence from quantum dots, an external magnetic field was required for two reasons (i) rotation of the magnetization of the electrode along the growth axis of the structure to establish clear optical selection rules for emission of circularly polarized photons \cite{Dyakonov:2008a}; (ii) Circularization of the eigenstates in the quantum dots that are linearly polarized due to their shape anisotropy \cite{Bayer:2002a}. Despite encouraging recent progress, an experimental demonstration of efficient electrical spin injection into a single quantum dot at zero magnetic field leading to highly circularly polarized electroluminescence is still lacking. \\ 
\begin{figure*}
\includegraphics[width=0.99\linewidth]{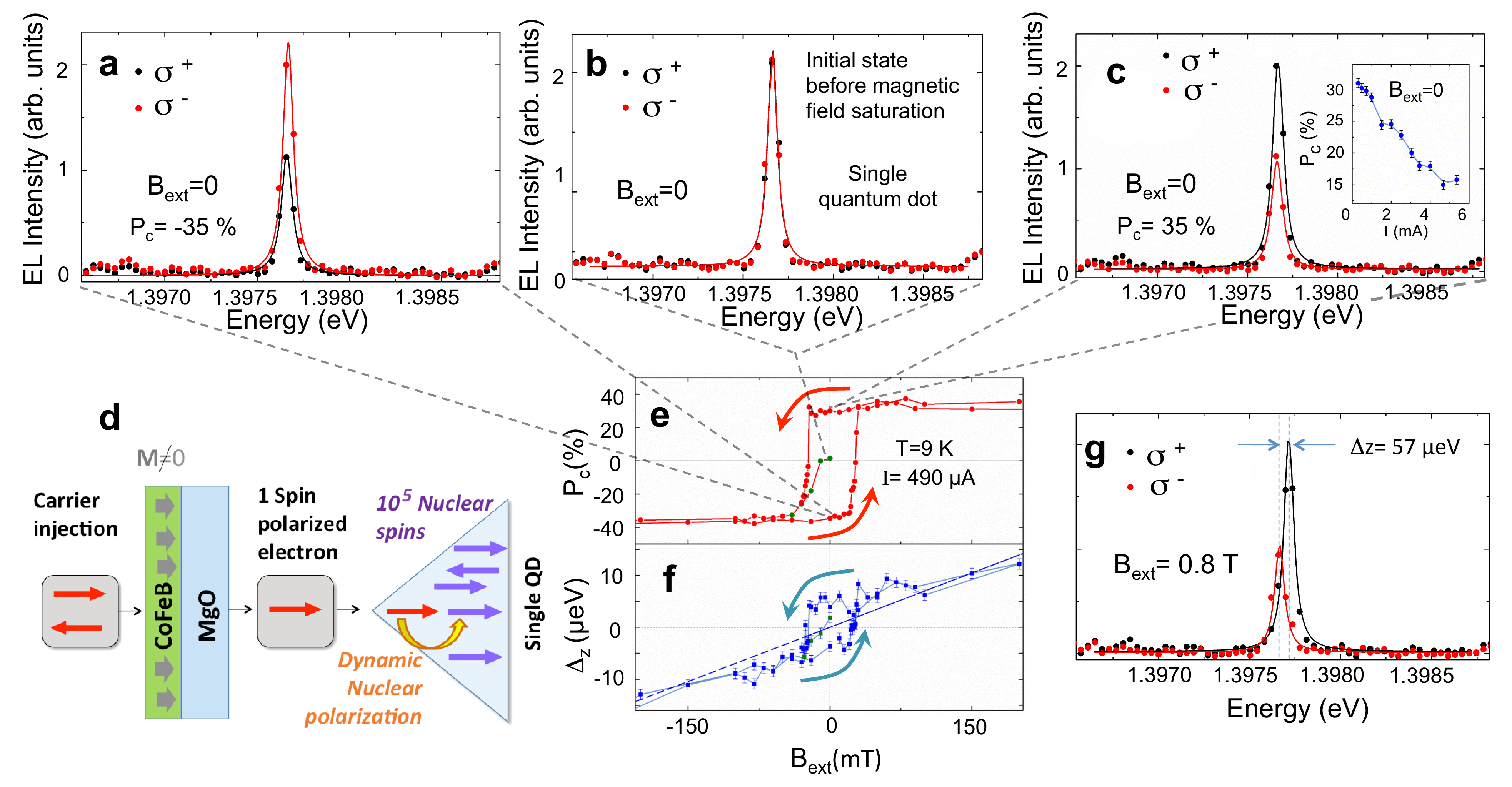}
\caption{\label{fig:fig2} \textbf{Polarization-resolved electroluminescence of a single quantum dot and electrical nuclear spin initialization. a. }Strongly polarized single dot emission at zero applied field (sweep direction negative to positive $B_{ext}$). The background emission has been subtracted for clarity. \textbf{b.} Unpolarized single dot emission before the first magnetization of the CoFeB layer \textbf{c.} Strongly polarized single dot emission at zero applied field (sweep direction positive to negative $B_{ext}$). The polarization degree $P_c$ is plotted as a function of the applied current in the inset. \textbf{d.} Schematics of principle of the electrical initialization of nuclear spins in a single quantum dot at zero magnetic field due to efficient electrical spin injection in the dot combined with the hyperfine interaction between electron and nuclear spins. \textbf{e.} The circular polarization degree of the Electroluminescence is measured as a function of the applied magnetic field $B_{ext}$. The sweep direction of the field is indicated by arrows. The data points corresponding to the initial magnetization are plotted in green. \textbf{f.} The energy difference $\Delta_Z=E^{EL}_{\sigma+}-E^{EL}_{\sigma-}$ is plotted as a function of $B_{ext}$. The dashed line corresponds to the linear fit of $\Delta_Z$ versus magnetic field for larger values of $B_{ext}$. \textbf{g.} Spectrally and polarization-resolved electroluminescence emission of a single dot at $B_{ext}=+0.8~$T.} 
\end{figure*} 
\indent Here we show that strongly circularly polarized electroluminescence emission from a single quantum dot can be achieved by combining two major device improvements. First, we use an ultrathin CoFeB injector with perpendicular magnetic anisotropy (PMA) \cite{Liang:2014a}, which is magnetized even at zero magnetic field after initial saturation. Second, instead of neutral quantum dots we use p-doped InAs/GaAs quantum dots (with one hole per dot on average) to benefit from simple optical selection rules for circularly polarized light of charged excitons (trions X$^+$) \cite{Marcinkevicius:2006a,Braun:2005a,Xu:2007a}. We demonstrate that electrical spin injection and optical read out of the average electron spin (at least 35\% spin polarization) is possible in a single quantum dot at zero magnetic field. Furthermore, the highly efficient electrical electron spin injection results in spin polarization of the nuclei of the atoms that form the dot mediated by the electron-nuclei hyperfine interaction. Nuclear spin polarization is reversed as we change the spin-orientation of the electrically injected electron spin. Controlling the nuclear spin bath in a dot in our Spin-LED is not just beneficial in order to potentially prolong carrier spin life and coherence times \cite{Urbaszek:2013a,Bechtold:2015a,Gangloff:2017a}, but the nuclear spins themselves with long relaxation times are an interesting system for memory applications \cite{Maletinsky:2009a,Waeber:2016a,Greilich:2007a}.\\
\indent \textbf{Polarization resolved electroluminescence of a quantum dot ensemble.--- } A schematic of the Spin-LED structure is shown in Figure 1b. The spin injector consist of a 2.5~nm thick MgO layer and a 1.1 nm Co$_{0.4}$Fe$_{0.4}$B$_{0.2}$ layer covered by a 5~nm Ta protection layer deposited by sputtering. The p-i-n quantum dot (QD) LED device grown by MBE contains a single layer of In$_{0.3}$Ga$_{0.7}$As QDs embedded in the optically active region with Be delta doping (p-type) near the quantum dot layer (see methods).  This favors the formation of the positively charged exciton X$^+$ (2 valence holes, 1 conduction electron). The X$^+$ consists of a hole spin singlet and an unpaired electron spin so the measured circular polarization $P_c$ in electroluminescence is directly given by the electron spin polarization as $P_c=-2 \langle S_e^z \rangle$ where $\langle S_e^z \rangle$  is the average electron spin projection onto the quantization axis (here also growth direction) \cite{Braun:2005a}. The measured EL degree of circular polarization is defined as $P_c=(I_{\sigma+}-I_{\sigma-})/(I_{\sigma+}+I_{\sigma-})$. Here $I_{\sigma+}$ (resp.  $I_{\sigma-}$ ) represents the integrated emission intensity of the right (left) circularly polarized EL component. The thin CoFeB/MgO spin injector possesses a strong perpendicular-magnetic-anisotropy (PMA) due to the interfacial anisotropy at the FM/Oxide interface \cite{Ikeda:2010a}. Once its magnetization is saturated through the application of a small out-of-plane magnetic field $B_{ext}$, this material retains a remnant out-of-plane magnetization even if the external field is switched off. This allows studying the spin-LED device with a magnetic electrode but at zero external applied field. The single dot electroluminescence (EL) is recorded at low temperature T=9~K in Faraday geometry, i.e. the magnetic field is applied along the growth axis, with a homebuilt confocal microscope with a detection spot diameter of  about $1~\mu$m  \cite{Vidal:2016a,Sallen:2014a}, see methods.\\  
\indent Figure 1d displays the EL spectra measured at zero magnetic field, which shows an ensemble of many spectrally narrow emission peaks, centered at about 1.36~eV, corresponding to emission of a large number of QDs. The QD ensemble EL emission is strongly circularly polarized with a large value of $P_c = 23\%$, although the applied magnetic field is zero. Here a DC voltage is used for LED operation, the generated current is sufficiently low (490~$\mu$A) to avoid any major impact of heating on the measurements. In Figure 1e we plot the degree of circular polarization $P_c$ of the ensemble EL of Figure 1d as a function of the applied magnetic field $B_{ext}$. We find for $P_c$ clear hysteresis behavior as a function of magnetic field sweep direction. This is in good agreement with the hysteresis loop of the normalized magnetization of CoFeB electrode measured by Superconducting Quantum Interference Device (SQUID) at T=30~K on an unpatterned sample \cite{Liang:2014a} (the slight discrepancy between the two curves is due to the difference in temperatures : 30~K instead of 9~K). This gives a strong indication that the EL polarization of the quantum dot emission is a reliable measure of the electron spin polarization, which in turn is determined by the out-of-plane magnetization of the FM electrode. Our device shows an EL circular polarization degree $P_c > 10\%$ at zero field up to liquid nitrogen temperatures, as demonstrated in the inset of Figure 1e. In previous studies application of magnetic fields of several Tesla was necessary to obtain strongly polarized EL emission from a single quantum dot \cite{Asshoff:2011a}. Here we achieve this goal in the absence of applied magnetic fields. \\
\indent \textbf{Polarization resolved electroluminescence on a single quantum dot.---} By using spectral filtering we are able to isolate the emission from a single quantum dot. In Figure 2a we show strongly circularly polarized emission with $P_c = -35\%$ at $B_{ext} = 0$ when the magnetic field is swept from negative to positive values. In contrast, when we sweep the field from positive to negative values, we record at $B_{ext} = 0$ a polarization of $P_c = +35\%$ in Figure 2c. This strong EL polarization following injection of spin-polarized electrons is a strong indication that the emission stems from the X$^+$ trion \cite{Bracker:2005a}. In addition, the absence of a clear doublet structure, as commonly observed for neutral exciton emission \cite{Bayer:2002a,Gammon:1996a} is another typical feature of trion emission. For positively charged excitons, optical selection rules yield  $P_c=-2 \langle S_e^z \rangle$.  Therefore, the results in Figure 2a and 2c indicate that the initially injected electron spin polarization is at least partially conserved during the radiative lifetime of $\approx 800~ps$ \cite{Braun:2005a,Paillard2001}(see discussion of nuclear spin fluctuations on electron spin decay below). In the inset of Figure 2c, it is shown that the EL circular polarization decreases with increasing current. This could be due to an increase of the kinetic energy of injected electrons with applied bias so that electron spin relaxation via the Dyakonov-Perel mechanism becomes more efficient \cite{Barate:2017a}.\\
\indent \textbf{Hysteresis cycle of the electroluminescence polarization of a single QD.---} The measurements at $B_{ext} = 0$ that result in strongly polarized EL rely on the PMA of the CoFeB electrode. Now we want to study the EL emission of a single dot as a function of the applied magnetic field $B_{ext}$ in more detail to check if the reversal of magnetization also results in injection of electrons with the opposite spin and hence changes the sign of $P_c$. At zero field, at the beginning of the measurements, the EL polarization is zero as shown in Figure 2b (see also the green points in Figure 2e). The domains in the electrode are initially randomly magnetized (up and down), which results on average in zero magnetization for the CoFeB layer. When applying an external out-of-plane magnetic field, the domains are gradually magnetized along this field, resulting in an increase of the average spin of the electrons $\langle S_e^z \rangle$ injected into the quantum dot and therefore of the polarization of the emitted light in EL. Once the magnetization is saturated, the CoFeB electrode presents a remnant magnetization when going back to zero magnetic field. Similar to the ensemble dot EL in Figure 1e, the circular polarization of the single dot EL changes sign at the critical fields around $\pm30~$mT due to the switching of the CoFeB magnetization. The main difference between Figure 1e and Figure 2e is the overall polarization degree, reaching above 35\% for the single dot compared to about 20\% for the ensemble dot EL. The results in Figure 2e show the direct link between the observed EL polarization and the average electron spin for X$^+$ emission as  $P_c=-2 \langle S_e^z \rangle$. \\
\indent \textbf{Zeeman splitting and nuclear spin polarization.} More surprising are the results shown in Figure 2f: here we plot the peak energy difference  $\Delta_z=E^{EL}_{\sigma+}-E^{EL}_{\sigma-}$ between the $\sigma^+$ and $\sigma^-$ polarized EL components as a function of the applied magnetic field. Whereas at $B_{ext} > 100$~mT the splitting $\Delta_z$  is a linear function of $B_{ext}$ as expected (not shown), the low and zero field data shown here exhibit again strong hysteresis, with the same coercivity as that of the EL circular polarization. $\Delta_z$ is measured to be up to $7~\mu$eV at  $B_{ext}=0$. Assuming an electron g-factor for a typical InGaAs dot of $|g|  = 0.6$ in $\Delta_z(0~\text{T})=g_e \mu_B B_n$ and neglecting the hole-nuclear spin interaction \cite{Urbaszek:2013a} this corresponds to an effective magnetic field of the order of $B_n\approx 200~$ mT experienced by the electron in the dot. The amplitude of the splitting measured in EL is the signature of dynamic nuclear spin polarization \cite{Strand:2005a,Gammon:1997a,Chekhovich:2012a}. All nuclei in InGaAs carry a nuclear spin \cite{Chekhovich:2012a}  and the hyperfine interaction between carrier and nuclear spins has been shown to be very efficient in III-V quantum dots \cite{Gammon:1997a,Tartakovskii:2007a,Urbaszek:2013a,Bluhm:2010b}. During the EL experiment, electrons with well oriented spin are injected into the investigated dot. This electron spin polarization can be transferred in part to the nuclear spin ensemble to create a non-zero average nuclear spin polarization $\langle I_n^z \rangle$. This is seen by the electrons as an effective magnetic field $B_n$ and results in a measurable Zeeman splitting even at \textit{zero} applied field (Overhauser shift), as clearly demonstrated before in optical orientation experiments on single III-V dots \cite{Lai:2006a,Sallen:2014a}. As the electron spin changes its sign, also the effective nuclear field changes its direction and therefore the measured splitting $\Delta_z$  changes its sign in the experiment in Figure 2f. The nuclear spin polarization is therefore most likely at the origin of the hysteretic behavior that we find for the splitting $\Delta_z$ around zero magnetic field. An effective magnetic field $B_n$ of about $200~$ mT will in part screen the nuclear field fluctuations $\delta B_n$, which are at the origin of electron spin relaxation and decoherence \cite{Merkulov:2002a,Braun:2005a,Bluhm:2010b,Bechtold:2015a}. The effective magnetic field $B_n$ due to nuclear spin polarization therefore helps to stabilize the electron spin polarization and thus to obtain highly circularly polarized EL \cite{Krebs:2008a}. At higher applied fields the Zeeman splitting due to the external field dominates the Overhauser shift. \\
\indent In order to ascribe the observed splitting in EL at zero external magnetic field to nuclear spin effects, we need to exclude the impact of possible stray magnetic fields on the single dot EL, we show that these stray fields are negligible in the following: The stray field is well known to be large (close to $\mu_0 M_S$, where $M_s$ is the magnetization at saturation) inside the ferromagnet due to the shape anisotropy and surface layer roughness does not change this value significantly. However, outside the ferromagnet but inside the semiconducting part of the device, the stray field is strictly zero in the absence of roughness, and negligibly small in a realistic situation. The roughness can be characterized by a modulation period of average amplitude $\sigma$ and lateral correlation length $\xi$ \cite{Bruno:1988a}. Considering the first order frequency contribution of the magnetic roughness (corresponding to a pure sinus shape roughness), the maximal value of the stray field can be written as 
$B_{max} = (\mu_0 M_s/\pi)[1-\text{exp}(-2 k_{00}\sigma)] \text{exp} (-k_{00} t)$, where $k_{00}=\sqrt{2}\pi/\xi$ is the characteristic wavevector frequency along the two in-plane directions and $t$ is the distance from the bottom surface of magnetic layer to the quantum dot layer. From AFM measurements performed on the MgO/GaAs system, we have estimated the average surface roughness amplitude $\sigma \approx 0.3~nm$ and the correlation length $\xi \approx50~$nm. Taking $M_s = 6.35 \times 10^5$~A/m for a CoFeB layer thickness of $1.1~$nm \cite{Liang:2014a}, and  $t=95~$nm leads to an estimation of the stray field of around  $B_{max}\approx 2.6~\mu$T. This is five orders of magnitude smaller than the effective field $B_n$ of roughly 200~mT extracted from our EL measurements and can therefore not explain the observed splitting $\Delta_z$ at $B_{ext}=0$ in Figure 2f. Note that a characteristic noise function introduced in the roughness function would even lead to a smaller value of the stray field. Therefore, we can rule out the possibility that the stray field is at the origin of the effective magnetic field felt by the electrons in a quantum dot. In addition, we can see that the energy splitting is close to zero in the initial state when the injector is not magnetized (see green points in fig. 1e). This is another important argument for dynamic nuclear polarization being responsible for the observed splitting in EL emission from an individual dot in the absence of any external magnetic field.\\
\indent \textbf{Conclusions.---}  We report strongly circularly polarized electroluminescence of a single InGaAs quantum dot in GaAs using an ultrathin CoFeB electrode. A polarization degree of up to 35\% (20\%) is observed for individual dots (dot ensembles) in the absence of applied external magnetic fields. This demonstrates that very efficient electrical spin injection and optical read-out of spin polarized electrons are possible in a single quantum dot without the need of an external magnetic field. Due to the efficient hyperfine interaction in III-V nanostructures, the repeated injection of spin polarized electrons into the dot leads to dynamic nuclear polarization and hence a measurable Overhauser shift. This paves the way for highly circularly polarized compact light sources based on ensembles or single quantum dots, as well as electrical initialization of a single quantum bit carried by the electron spin, or alternatively by the nuclear spin ensemble in a single dot. \\
\indent \textbf{Methods.---}\\
\textit{Sample growth: } The p-i-n LED device grown by MBE contains a single layer of In$ _{0.3}$Ga$ _{0.7}$As quantum dots embedded in the active region. The full sequence of the structure is the following: $p^+$-GaAs$:$Zn (001) substrate ($p = 3 \times 10^{18} cm^{-3} )$
 $|$ 300~nm p-GaAs:Be (p = 5$\times 10^{18}$~cm$^{-3})$ 
 $|$400~nm p-Al$ _{0.3}$Ga$_{0.7}$As:Be (p = 5$\times 10^{17} - 5 \times 10^{18}$~cm$^{-3}$) 
 $|$30~nm GaAs with Be delta doping near the QD layer and 1 layer of InGaAs cone-shaped quantum dots (density $1.6\times 10^{14}$~m$^{-2}$, bottom diameter 30~nm, height 9~nm) $|30~$nm intrinsic GaAs $|$50 nm n-GaAs$:$Si (n = 10$^{16}$~cm$^{-3}$). 
The LED was passivated with arsenic in the MBE chamber. Then, the structure was transferred through air into a second MBE-sputtering interconnected system. The As capping layer is firstly desorbed at 300 deg C in the MBE chamber and then the sample was transferred through ultra-high vacuum to a sputtering chamber to grow the MgO layer of thickness 2.5 nm. Finally, a 1.1 nm thick Co$_{0.4}$Fe$_{0.4}$B$_{0.2}$ spin injector and a 5~nm thick Ta protection layer are deposited by sputtering in both cases. Concerning the device fabrication, 300~$\mu$m diameter circular mesas were then processed using standard UV photolithography and etching techniques. Finally, the processed wafers were cut into small pieces to perform rapid temperature annealing (RTA) at 300  deg C for 3 minutes. More details of growth and optimization of the perpendicular spin-injector can be found in \cite{Tao:2016a,Liang:2014a}. \\
\textit{Transmission electron microscopy measurements: }High-resolution transmission electron microscopy (HR-TEM) studies were performed by using a JEOL ARM200 cold field-emission gun working at 200~kV.\\
\textit{Optical characterization: }The single dot electroluminescence (EL) is recorded in Faraday geometry with a home build confocal microscope with a detection spot diameter of ~1$\mu$m \cite{Vidal:2016a,Sallen:2014a}. The detected EL signal is dispersed by a spectrometer with 1200 grooves per mm and detected by cooled a Si-CCD camera with the spectral precision of 2~$\mu$eV. The polarization analysis  of the EL emission is performed with polarizers and achromatic wave-plates. Magnetic fields $B_{ext}$ perpendicular to the LED are generated by a superconducting coil inside a vibration-free closed cycle Helium cryostat, where the sample is mounted on nano-positioners.\\
\textit{AFM characterization:} Samples are scanned by a Solver P47 system (NT-MDT) with regular CONTACT mode.\\
\indent \textbf{Acknowledgments.---}
    We acknowledge funding from ERC Grant No. 306719. F.C., P.R. and H.C. acknowledge the grant Next No ANR-10 LABX-0037 in the framework of the "Programme des Investissements d'Avenir". X.M. acknowledges Institut Universitaire de France. Y.L acknowledges the support by the joint French National Research Agency (ANR)-National Natural Science Foundation of China (NSFC) SISTER project (Grants No. ANR-11-IS10-0001 and No.NNSFC 61161130527) and ENSEMBLE project (Grants No. ANR-14-0028-01 and No. NNSFC 61411136001). A. D. acknowledges funding  by Region Lorraine.

 

\end{document}